\documentclass[twocolumn,journal]{IEEEtran}
\usepackage[T1]{fontenc}
\usepackage[latin9]{inputenc}
\usepackage{color}
\usepackage{textcomp}
\usepackage{amsbsy}
\usepackage{amstext}
\usepackage{amsmath}
\usepackage{graphicx}
\usepackage{dsfont}
\usepackage{amsfonts}
\usepackage{amssymb}
\usepackage[scr=rsfs]{mathalpha}

\usepackage[unicode=true,
 bookmarks=true,bookmarksnumbered=true,bookmarksopen=true,bookmarksopenlevel=1,
 breaklinks=false,pdfborder={0 0 0},pdfborderstyle={},backref=false,colorlinks=false]
 {hyperref}
\hypersetup{pdftitle={Time Modulation to Manage and Increase the Power Harvested From External Vibrations},
 pdfauthor={Kasra Rouhi},
 pdfpagelayout=OneColumn, pdfnewwindow=true, pdfstartview=XYZ, plainpages=false}

\usepackage{fancyhdr}

\makeatletter
\let\oldforeign@language\foreign@language
\DeclareRobustCommand{\foreign@language}[1]{%
  \lowercase{\oldforeign@language{#1}}}

\usepackage[caption=false,font=footnotesize]{subfig}

\makeatother

\begin{document}
\title{Time Modulation to Manage and Increase the Power Harvested From External
Vibrations}
\author{Alireza Nikzamir{*}, Kasra Rouhi{*}, Alexander Figotin, and Filippo
Capolino\thanks{Alireza Nikzamir, Kasra Rouhi, and Filippo Capolino are with the Department of Electrical Engineering and Computer Science, University of California,
Irvine, CA 92697 USA, e-mails: \protect\href{mailto:anikzami@uci.edu}{anikzami@uci.edu},
\protect\href{mailto:kasra.rouhi@uci.edu}{kasra.rouhi@uci.edu} and
\protect\href{mailto:f.capolino@uci.edu}{f.capolino@uci.edu}.}\thanks{Alexander Figotin is with the Department of Mathematics, University of California, Irvine, CA 92697 USA, e-mail: \protect\href{mailto:afigotin@uci.edu}{afigotin@uci.edu}.}\thanks{{*} These authors contributed equally.}}

\maketitle

\thispagestyle{fancy}

\begin{abstract}
We investigate how a single resonator with a time-modulated component
extracts power from an external ambient source. However, the collected
power is largely dependent on the precise choice of the modulation
signal frequency. We focus on the power absorbed from external vibration
using a one degree-of-freedom mechanical resonator where the damper
has a time-varying component. We show that time modulation can make
a significant difference in the amount of harvested power, leading
to more than 10 times enhancement with respect to an analogous system
without time modulation. We also find that a narrow band pair of peak
and dip in the spectrum of the absorbed power occurs because of the
presence of an exceptional point of degeneracy (EPD). In this narrow
frequency range, the delay between the damper modulating signal and
the external vibrating signal largely affects the collected power.
The high frequency-selectivity of EPD-induced power management could
potentially be used in sensing and spectrometer applications.
\end{abstract}


\IEEEpeerreviewmaketitle{}

Energy harvesting has attracted considerable interest in recent years
in electrical \cite{priya2005modeling,wickenheiser2010power,motter2012vibration}
and mechanical systems \cite{stephen2006energy,anton2007review,kim2011broadband}.
Energy harvesting technology offers a battery-less strategy by recovering
energy from ambient sources such as vibrations, wind, light etc.,
and transform it into another form, such as electrical power. Applications
include micro-electromechanical systems (MEMS) vibration energy harvesters
\cite{naito2019electrostatic,toshiyoshi2019mems}, low-power wireless
sensors \cite{paradiso2005energy}, fluid energy harvesting \cite{kwon2010t},
and biomechanical energy harvesters \cite{li2009development}. In
some of these applications, only a small fraction of energy needs
to be extracted to power devices, and it is particularly important
when devices are isolated. Therefore, collected power from a nearby
ambient source can be used to power the circuits inside the device.
Focusing on mechanical mechanisms, different vibration-based energy
harvesting methods have been used recently \cite{arrieta2010piezoelectric,jung2010energy,zilletti2012optimisation,di2013energy,tehrani2014extending,scapolan2016energy},
though more work needs to be done on utilizing parametric excitation
in time-varying systems for vibration energy harvesting. Furthermore,
analogous principles could be used to manage the absorption of vibration
or filter out particular vibration frequencies in mechanical systems
\cite{jiao2018intermodal}.

In general, low vibration amplitudes cannot be efficiently collected.
Therefore, various mechanical and electrical approaches have been
developed to increase energy harvesting efficiency \cite{zhu2009strategies,tang2011enhanced,bibo2013investigation}.
Vibration energy harvesters have been proposed with different nonlinear
arrangements that increase their frequency range and dynamic range,
most notably using nonlinear springs and dampers \cite{ramlan2010potential,laalej2012application}.
Moreover, semi-active strategies and nonlinear damping in the form
of cubic damping \cite{di2013energy,tehrani2014extending} and nonlinear
piezoelectric converters \cite{ferrari2010improved,cao2015nonlinear}
have been used to extend an energy harvester's dynamic range. In addition,
there is a lot of work on analyzing systems where the mass changes
over time \cite{van2006free,van2010forced,zukovic2017insight}. However,
challenges remain about maximizing the amount of energy harvested,
and new ideas should be further explored.

\begin{figure}[t]
\centering{}\includegraphics[width=0.8\columnwidth]{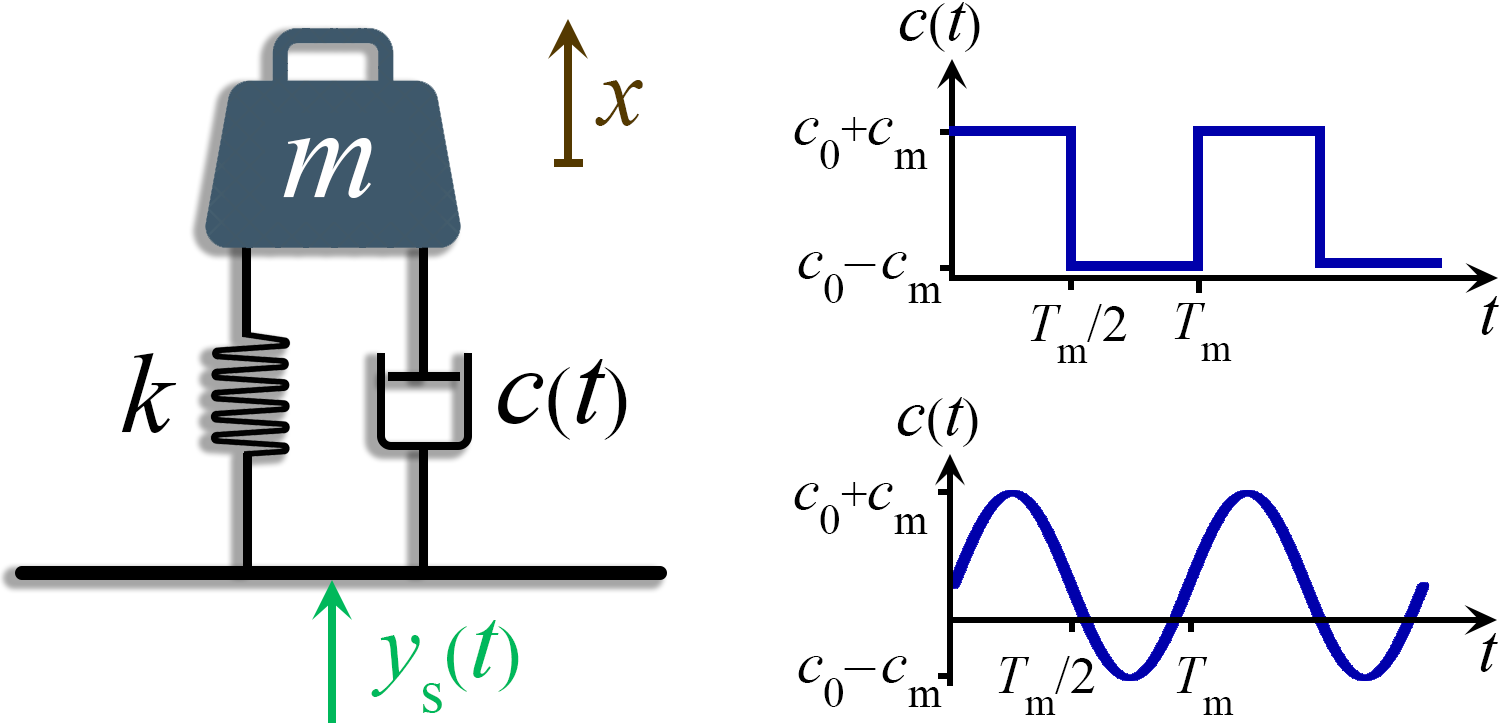}\caption{Time-modulated mechanical system for kinetic energy harvesting. The
external vibrational displacement of the whole system is $y_{\mathrm{s}}\left(t\right)$.
Two examples of time modulation of the damper are sinusoidal and two-level
piece-wise constant (used in this paper).\label{Fig: MechSystem}}
\end{figure}

In this letter, a resonator with a linear time-periodic (LTP) damper
is considered for harvesting or managing energy from an outside source,
focusing on a mechanical mass-spring-damper resonator subject to external
vibration \cite{stephen2006energy}, as shown in Fig. \ref{Fig: MechSystem}.
However, the physical principle here discussed is general and can
also be applied to other systems as shown in the Appendix
A. We demonstrate how parametric LTP modulation can boost the amplitude
of motion, enabling a more efficient flow from the energy source to
the harvesting system. By applying time variation to the system, power
harvesting is improved by 10 times in a specific frequency range compared
to the unmodulated system. Moreover, we observe extremely narrow spectral
features in the spectrum of the harvested energy and we provide an
explanation by resorting to the concept of exceptional points of degeneracy
(EPD). Such degeneracy is a point in the parameter space of a resonating
system at which multiple eigenmodes coalesce in both their eigenvalues
and eigenvectors \cite{heiss2004exceptional}. The presence of a nontrivial
Jordan block in the Jordan canonical form of the system matrix describing
the system's evolution shows the occurrence of EPD, as was demonstrated
in \cite{vishik1960solution,seyranian1993sensitivity,Figotin2003Oblique}.
The concept of EPD has been investigated in circuits with loss and/or
gain under parity-time symmetry \cite{stehmann2004observation,Schindler2011Experimental},
and also in spatially \cite{nada2017theory,abdelshafy2019exceptional}
and temporally periodic structures \cite{kazemi2019exceptional,rouhi2020exceptional,kazemi2022experimental}.
Moreover, the degenerate eigenvalues of the system are exceptionally
sensitive to any perturbations in system parameters \cite{Kato1966Perturbation}.
The EPD-based principle has been proposed to achieve high sensitivity
in various possible sensing scenarios, including optical microcavities
\cite{Hodaei2017Enhanced}, electron beam devices \cite{rouhi2021exceptional},
electronic resonators \cite{Schindler2011Experimental,Chen2018Generalized,rouhi2022exceptional,Nikzamir2022Highly},
and optomechanical mass sensors \cite{djorwe2019exceptional}.

The kinetic energy harvesting mechanical scheme discussed in this
letter is shown in Fig. \ref{Fig: MechSystem}. The system consists
of a mass connected to a spring and a damper with an additional time-varying
portion, and it is excited by external vibration represented by the
imposed displacement $y_{\mathrm{s}}\left(t\right)$ assumed to be
monochromatic. Here, $m$ is the mass, $k$ is the spring stiffness
constant, $c\left(t\right)=c_{0}+c_{\mathrm{m}}\left(t\right)$ is
the damping parameter that includes a constant part $c_{0}$ and a
time-periodic one $c_{\mathrm{m}}\left(t\right)$ of period $T_{\mathrm{m}}$
(see right panel in Fig. \ref{Fig: MechSystem}). Also, $x$ represents
the mass displacement, and $y_{\mathrm{s}}\left(t\right)$ is the
imposed displacement of the system caused by an external force that
drives the mechanical resonator. In addition, the electric counterpart
circuit is shown in Appendix A. The governing equation
of the time-varying system is

\begin{equation}
m\ddot{x}+c\left(t\right)\left(\dot{x}-\dot{y}_{\mathrm{s}}\right)+k\left(x-y_{\mathrm{s}}\right)=0.
\end{equation}

The constant damping coefficient $c_{0}=c_{\mathrm{p}}+c_{\mathrm{t}}$
represents the energy losses within the system due to parasitic loss
mechanisms $c_{\mathrm{p}}$ (e.g., viscous friction with air), and
by the intentional mechanism of damping $c_{\mathrm{t}}$, i.e., the
mechanical energy extracted by the transduction mechanism \cite{beeby2006energy,beeby2007micro,kundu2016modeling}.
Hence, part of the mobile mass\textquoteright s kinetic energy is
lost in mechanical parasitic damping and some other is turned into
electricity thanks to an energy converter (e.g., magnet/coil, piezoelectric
material, variable capacitor, etc.) \cite{erturk2009piezomagnetoelastic,boisseau2013nonlinear,basset2016electrostatic}.
Here, we presume that the mechanical damping force is proportional
to the velocity, which can be described as an electromechanical transducer
\cite[Ch. 2]{spies2015handbook}. It is assumed that the mass of the
vibration source is significantly greater than the mass $m$, so the
external vibration is just modeled as an imposed force. We define
the relative mass displacement parameter $z=x-y_{\mathrm{s}}$, and
the governing dynamic equation is rewritten as

\begin{equation}
\ddot{z}+2\zeta\left(t\right)\omega_{0}\dot{z}+\omega_{0}^{2}z=-\ddot{y}_{\mathrm{s}},\label{eq:diff_eq}
\end{equation}
where $\zeta\left(t\right)=\zeta_{0}+\zeta_{\mathrm{m}}\left(t\right)=c_{0}/\left(2m\omega_{0}\right)+c_{\mathrm{m}}\left(t\right)/\left(2m\omega_{0}\right)$
is the time-modulated damping rate and $\omega_{0}=\sqrt{k/m}$ is
the natural frequency of the unmodulated and lossless system. Assuming
a time harmonic dependence of the form $z\propto e^{j\omega t}$ for
the \textit{unmodulated} homogeneous system (i.e., $c_{\mathrm{m}}=0$
and $y_{\mathrm{s}}=0$), we obtain the complex eigenfrequencies $\omega=\omega_{0}\left(\pm\sqrt{1-\zeta_{0}^{2}}+j\zeta_{0}\right)$
associated to damped oscillations. To analyze the time-modulated system,
we define the state vector as $\boldsymbol{\Psi}\left(t\right)\equiv\left[z,\dot{z}\right]{}^{\mathrm{T}}$,
where the superscript $\mathrm{T}$ denotes the transpose operation,
leading to

\begin{equation}
\begin{array}{c}
\begin{array}{c}
\mathrm{d\boldsymbol{\Psi}\left(\mathit{t}\right)/d\mathit{t}=\mathbf{\underline{M}}\left(\mathit{t}\right)\boldsymbol{\Psi}\left(\mathit{t}\right)+\left(\begin{array}{c}
0\\
-\ddot{y}_{\mathrm{s}}
\end{array}\right)},\end{array}\\
\\
\mathbf{\underline{M}}\left(t\right)=\left(\begin{array}{cc}
0 & 1\\
-\omega_{0}^{2} & -2\zeta\left(t\right)\omega_{0}
\end{array}\right),
\end{array}\label{eq:Damper_Main}
\end{equation}
where $\mathbf{\underline{M}}\left(t\right)$ is the time-variant
system matrix. We first analyze the power transfer from an external
monochromatic vibration $y_{\mathrm{s}}\left(t\right)=y_{0}\cos\left(2\pi f_{\mathrm{s}}t\right)$,
where $y_{0}$ is its amplitude and $f_{\mathrm{s}}$ is its frequency,
into the LTP spring-mass-damper system using a time-domain numerical
simulator (see Appendix A). We determine the time-averaged
power $\mathrm{\mathit{P}_{s}}$ delivered by the external source
$y_{\mathrm{s}}\left(t\right)$, the time-averaged power $\mathrm{\mathit{P}_{m}}$
delivered by the time modulation, and the time-averaged power $P_{0}$
delivered to (or harvested by) the constant damper $c_{0}$. Note
that $f_{\mathrm{m}}$ is the modulation frequency of the time-varying
damper, $f_{\mathrm{s}}$ is the frequency of the external source.
In some designs, ambient vibrations are generally low in amplitude
and frequency \cite{boisseau2013nonlinear}, so, we investigate how
we can maximize both the power $\mathrm{\mathit{P}_{s}}$ absorbed
from the external source and the power $P_{0}$ harvested by the constant
damper by using time modulation.

\begin{figure*}[tbh]
\centering{}\includegraphics[width=0.85\textwidth]{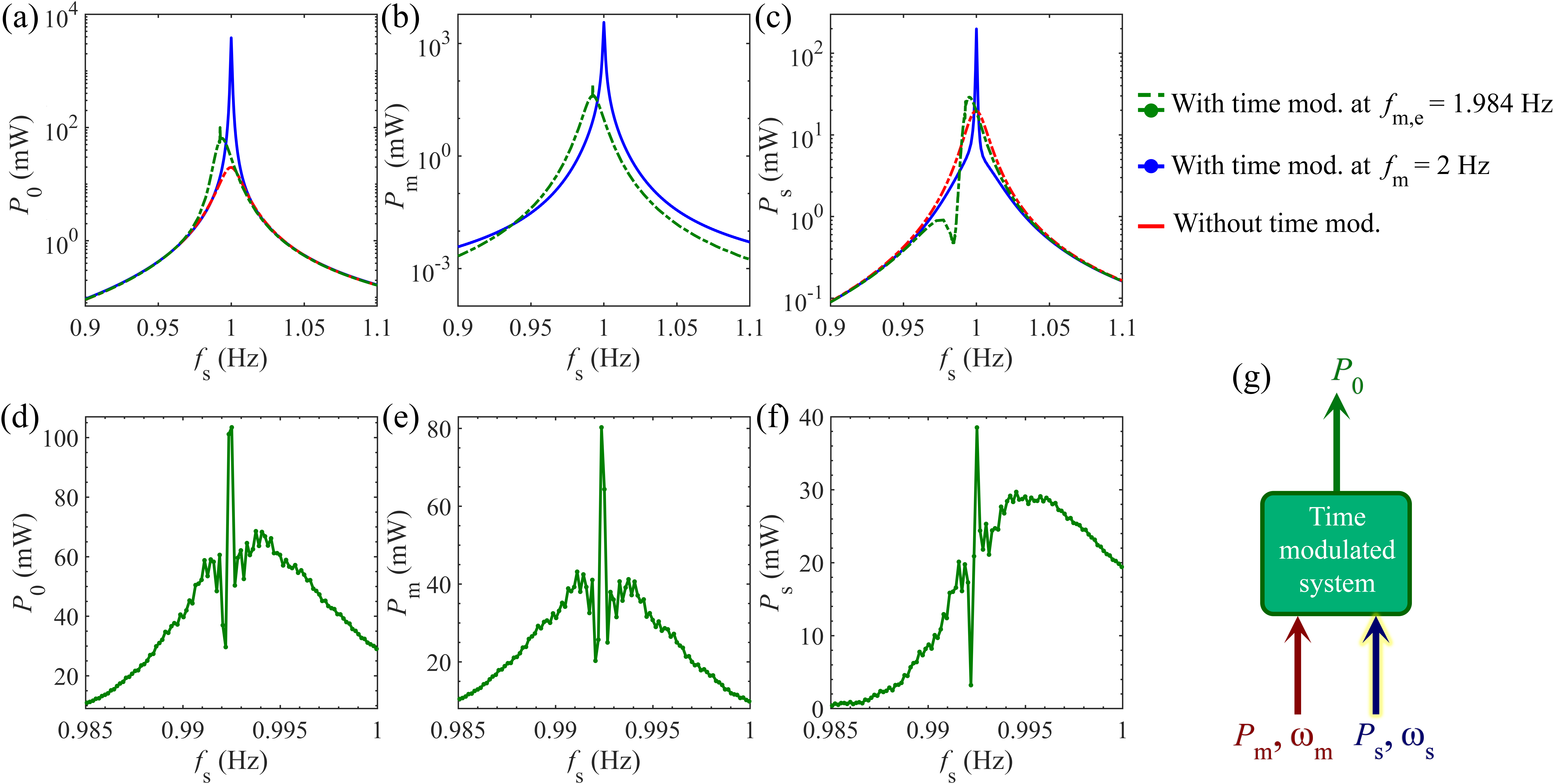}\caption{Time-averaged power levels after reaching steady state for: (a) $P_{0}$
delivered to the damper $c_{0}$; (b) $P_{\mathrm{m}}$ provided by
the time-varying damper $c_{\mathrm{m}}\left(t\right)$; and (c) $P_{\mathrm{s}}$
extracted (harvested) from the external vibration, by varying the
vibration frequency $f_{\mathrm{s}}$ in a wide range around $f_{\mathrm{0}}=1\:\mathrm{Hz}$.
Two time-modulated cases are considered here: modulation frequency
$f_{\mathrm{m}}=2\:\mathrm{Hz}$ at the center of the modulation gap
(blue), and modulation frequency $f_{\mathrm{m}}=f_{\mathrm{m,e}}=1.984\:\mathrm{Hz}$
(green, note the sharp feature). For comparison, powers are also shown
for the case without time modulation (red). (d)-(f) Zoomed-in analysis
for frequencies near the EPD frequency $f_{\mathrm{e}}=f_{\mathrm{m,e}}/2=0.992\:\mathrm{Hz}$
for the case with time modulation at $f_{\mathrm{m}}=f_{\mathrm{m,e}}=1.984\:\mathrm{Hz}$.
There is a remarkable highly varying power level around $f_{\mathrm{e}}$.
(g) Schematic of power diagram showing that the collected time-average
power $P_{0}$ equals the sum of $P_{\mathrm{m}}$ and $P_{\mathrm{s}}$.\label{Fig:excitation_fs}}
\end{figure*}

A specific example is shown in Fig. \ref{Fig:excitation_fs}, where
$k=4\pi^{2}\:\mathrm{N/m}$ and $m=1\:\mathrm{kg}$ leading to $f_{\mathrm{0}}=1\:\mathrm{Hz}$,
and $y_{0}=1\:\mathrm{mm}$. We only consider for simplicity a two-level
piece-wise constant time-periodic damping $c\left(t\right)$, which
is equal to $c_{0}+c_{\mathrm{m}}$ in the time interval $0\leqslant t<T_{\mathrm{m}}/2$
and equal to $c_{0}-c_{\mathrm{m}}$ in $T_{\mathrm{m}}/2\leqslant t<T_{\mathrm{m}}$.
In this example, we assume $c_{0}=0.1\:\mathrm{Ns/m}$ and $c_{\mathrm{m}}=0.15\:\mathrm{Ns/m}$.
We also study the unmodulated system with constant damper $c\left(t\right)=c_{0}$,
where maximum energy can be extracted when the excitation frequency
$f_{\mathrm{s}}$ matches the natural frequency of the system $f_{0}$,
as explained in Appendix B. In Fig. \ref{Fig:excitation_fs}
we compare the LTP system with the unmodulated system to show that
time modulation has a strong effect on the time-averaged power $\mathrm{\mathit{P}_{s}}$
absorbed from the source and on the power delivered to the constant
portion of the damper $P_{0}$. For the considered modulation frequency
of $f_{\mathrm{m}}=2\:\mathrm{Hz}$ that is equal to $2f_{0}$, $P_{\mathrm{s}}$
and $P_{0}$ are largely enhanced when $f_{\mathrm{s}}$ is in the
neighbor of $f_{0}$. The plot in Fig. \ref{Fig:excitation_fs}(c)
shows the maximum harvested power in the unmodulated system is $P_{\mathrm{s}}=19.7\:\mathrm{mW}$,
whereas the maximum power that the time modulated system absorbs from
the source is $P_{\mathrm{s}}=198.4\:\mathrm{mW}$, i.e., 10 times
higher than that of the unmodulated system.

The results in Figs. \ref{Fig:excitation_fs}(a)-(c) show also another
interesting feature, i.e., the very narrow frequency range around
$f_{\mathrm{s}}=0.992\:\mathrm{Hz}$ where the power exhibits a sharp
peak with abrupt changes when $f_{\mathrm{m}}=f_{\mathrm{m,e}}$,
where $f_{\mathrm{m,e}}=1.984\:\mathrm{Hz}$ is a modulation frequency
that leads to the EPD, as shown next. This rapid power level variation
is shown better in the zoomed-in frequency region in Figs. \ref{Fig:excitation_fs}(d)-(f).
The power level experiences a sharp maximum and a local minimum near
$f_{\mathrm{s}}=0.992\:\mathrm{Hz}$ (which is half of $f_{\mathrm{m,e}}$).
To understand the reasons for this very sharp peak and large variations
in the time averaged power values we look at the eigenvalues of the
system and their degeneracy.

To justify some of the above results, we look at the eigenstates
of the time-varying system in the absence of an external vibration
source, i.e., when $y_{\mathrm{s}}=0$. The evolution of the state
vector in the LTP system with time periodicity $T_{\mathrm{m}}$,
from the time $t$ to $t+T_{\mathrm{m}}$, is given by $\boldsymbol{\Psi}\left(t+T_{\mathrm{m}}\right)=\boldsymbol{\Phi}\left(t+T_{\mathrm{m}},t\right)\boldsymbol{\Psi}\left(t\right)$,
where $\boldsymbol{\Phi}\left(t+T_{\mathrm{m}},t\right)$ is the state
transition matrix, which is determined as $\underline{\boldsymbol{\Phi}}=e^{\mathbf{\underline{M}}_{2}T_{\mathrm{m}}/2}e^{\mathbf{\underline{M}}_{1}T_{\mathrm{m}}/2}$,
where $\mathbf{\underline{M}}_{1}$ and $\mathbf{\underline{M}}_{2}$
are the system matrices in the first and second time intervals \cite[Ch. 2]{richards2012analysis}.
We look for eigensolutions of the system that satisfy $\boldsymbol{\Psi}\left(t+T_{\mathrm{m}}\right)=e^{-j\omega T_{\mathrm{m}}}\boldsymbol{\Psi}\left(t\right)$
where $\omega$ (with all the harmonics $\omega+2\pi q/T_{\mathrm{m}}$,
where $q$ is an integer) is the complex eigenfrequency. Therefore,
the eigenvalue problem is

\begin{figure}[t]
\centering{}\includegraphics[width=1\columnwidth]{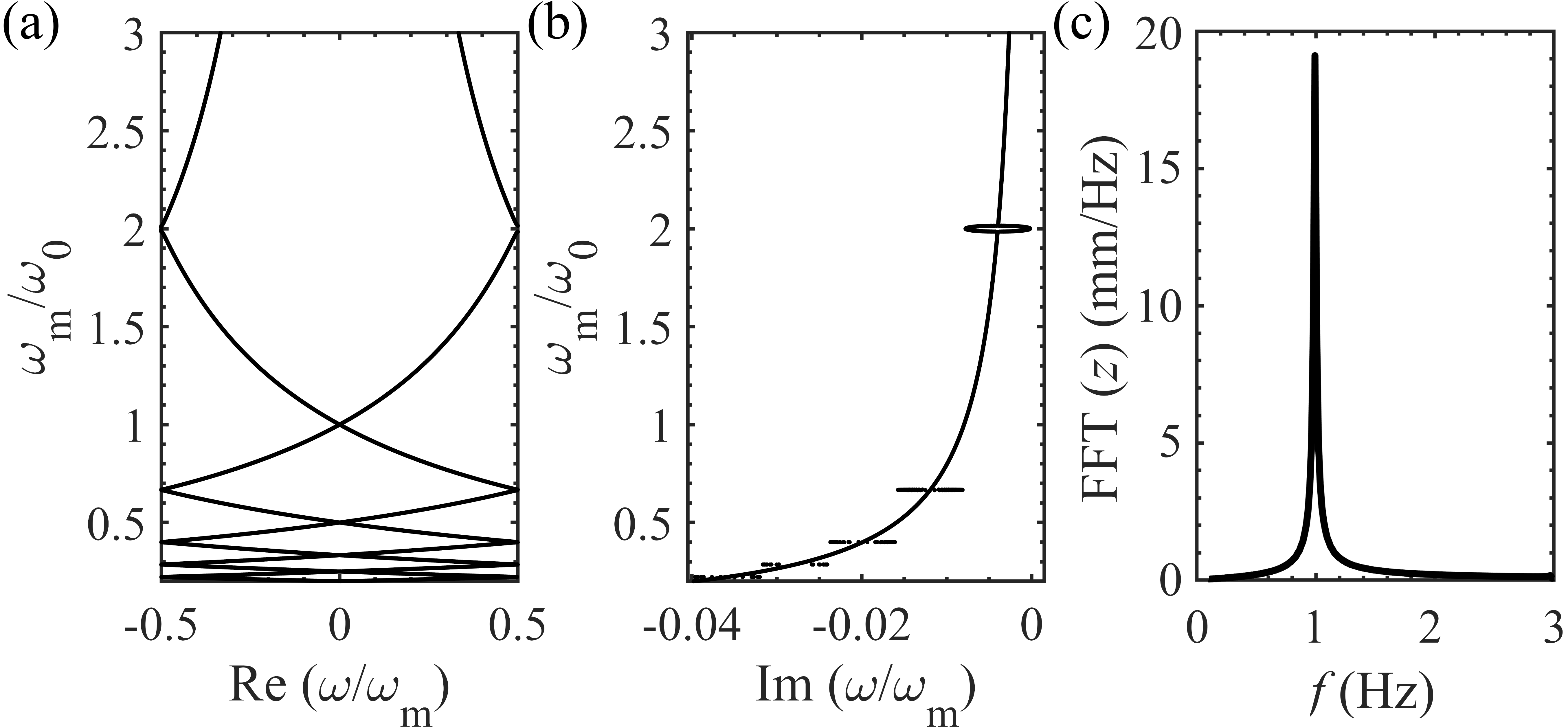}\caption{The (a) real and (b) imaginary parts of eigenfrequencies $\omega+q\omega_{\mathrm{m}}$,
where $q$ is an integer, of the system by varying $\omega_{\mathrm{m}}$.
(c) Frequency spectrum of the relative mass displacement $z\left(t\right)$.
The largest frequency spectral component of the displacement occurs
at the fundamental harmonic $q=0$, i.e., at the EPD frequency $f_{\mathrm{e}}=f_{\mathrm{m,e}}/2=0.992\:\mathrm{Hz}$.\label{Fig: Dispersion_diagram}}
\end{figure}

\begin{equation}
\boldsymbol{\underline{\Phi}}\boldsymbol{\Psi}\left(t\right)=\lambda\boldsymbol{\Psi}\left(t\right),\label{eq: EigVal}
\end{equation}
and the eigenvalues $\lambda_{n}=e{}^{-j\omega_{n}T_{\mathrm{m}}},n=1,2$,
are obtained by solving the characteristic polynomial equation $\mathrm{det}\left(\underline{\boldsymbol{\Phi}}-\lambda\mathrm{\underline{\boldsymbol{I}}}\right)=0$.
The eigensolutions $\boldsymbol{\Psi}\left(t\right)$ have Fourier
harmonics with frequencies $\omega_{n}+q\omega_{\mathrm{m}}$, where
$\omega_{\mathrm{m}}=2\pi f_{\mathrm{m}}$, is the modulation angular
frequency \cite{kazemi2019exceptional}. When a transition matrix
made of real values elements describes the system, the characteristic
polynomial has real coefficients, so the eigenvalues are either real
or complex conjugate pairs. The determinant of the transition matrix
is written as \cite[Ch. 2]{richards2012analysis}

\begin{equation}
\mathrm{det}\left(\underline{\boldsymbol{\Phi}}\right)=\lambda_{1}\lambda_{2}=e^{\left[\mathrm{tr}\left(\mathbf{\underline{M}}_{1}T_{\mathrm{m}}/2\right)+\mathrm{tr}\left(\mathbf{\underline{M}}_{2}T_{\mathrm{m}}/2\right)\right]},
\end{equation}
where tr is the trace of the matrix. The determinant can be either
$\mathrm{det}\left(\underline{\boldsymbol{\Phi}}\right)=e^{2\mathrm{Im}\left(\omega_{1}\right)T_{\mathrm{m}}}$,
when eigenvalues are complex conjugate pair, i.e., $\lambda_{2}=\lambda_{1}^{*}$,
or $\mathrm{det}\left(\underline{\boldsymbol{\Phi}}\right)=e^{js\pi}e^{\left(\mathrm{Im}\left(\omega_{1}\right)+\mathrm{Im}\left(\omega_{2}\right)\right)T_{\mathrm{m}}}$,
when $\lambda_{1}$ and $\lambda_{2}$ are both real and $s$ is an
integer. The two eigenvalues are

\begin{equation}
\lambda_{1,2}=\frac{\mathrm{tr}\left(\underline{\boldsymbol{\Phi}}\right)}{2}\pm\sqrt{\left(\frac{\mathrm{tr}\left(\underline{\boldsymbol{\Phi}}\right)}{2}\right)^{2}-\mathrm{det}\left(\underline{\boldsymbol{\Phi}}\right)},\label{eq: GenEigVal}
\end{equation}
and the two associated eigenvectors are

\begin{equation}
\boldsymbol{\Psi}_{1}=\left[\begin{array}{c}
\varphi_{12}\\
\lambda_{1}-\varphi_{11}
\end{array}\right],\;\boldsymbol{\Psi}_{2}=\left[\begin{array}{c}
\varphi_{12}\\
\lambda_{2}-\varphi_{11}
\end{array}\right],\label{eq: GenEigVec}
\end{equation}
where $\varphi_{11}$ and $\varphi_{12}$ are elements of the matrix
$\underline{\boldsymbol{\Phi}}$. The two eigenvalues are degenerate
($\lambda_{1}=\lambda_{2}=\mathrm{tr}\left(\underline{\boldsymbol{\Phi}}\right)/2$)
when

\begin{equation}
\mathrm{tr}\left(\underline{\boldsymbol{\Phi}}\right)=\pm2\sqrt{\mathrm{det}\left(\underline{\boldsymbol{\Phi}}\right)}.\label{eq: EPDCondition}
\end{equation}

\begin{figure*}[t]
	\centering{}\includegraphics[width=0.8\textwidth]{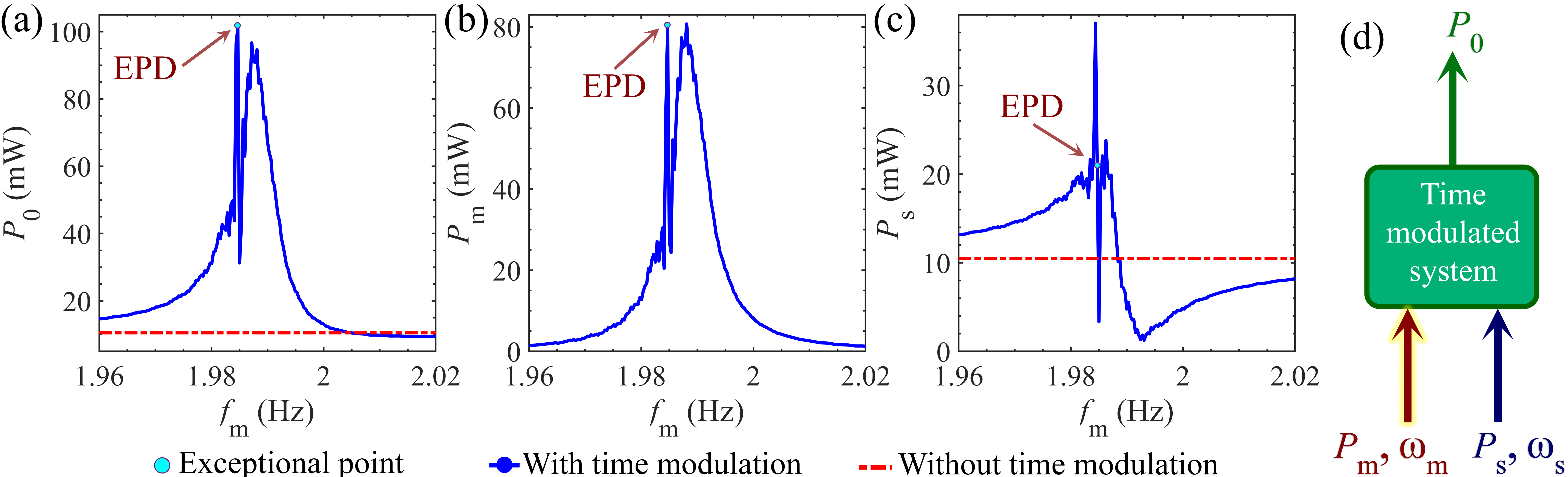}\caption{Time-averaged powers by varying the modulation frequency $f_{\mathrm{m}}$
		(blue curves), for the case of $f_{\mathrm{s}}=f_{\mathrm{m,e}}/2$.
		(a) $P_{0}$ delivered to the constant damper $c_{0}$; (b) $P_{\mathrm{m}}$
		delivered by the time-varying damper $c_{\mathrm{m}}\left(t\right)$;
		and (c) $P_{\mathrm{s}}$ collected from the source. The red dashed
		line is a reminder of the time-averaged power level of the unmodulated
		system. (d) Schematic of power diagram showing that the collected
		time-average power $P_{0}$ equals the sum of $P_{\mathrm{m}}$ and
		$P_{\mathrm{s}}$.\label{Fig:modulation_fm}}
\end{figure*}

\begin{figure}[tbh]
	\centering{}\includegraphics[width=1\columnwidth]{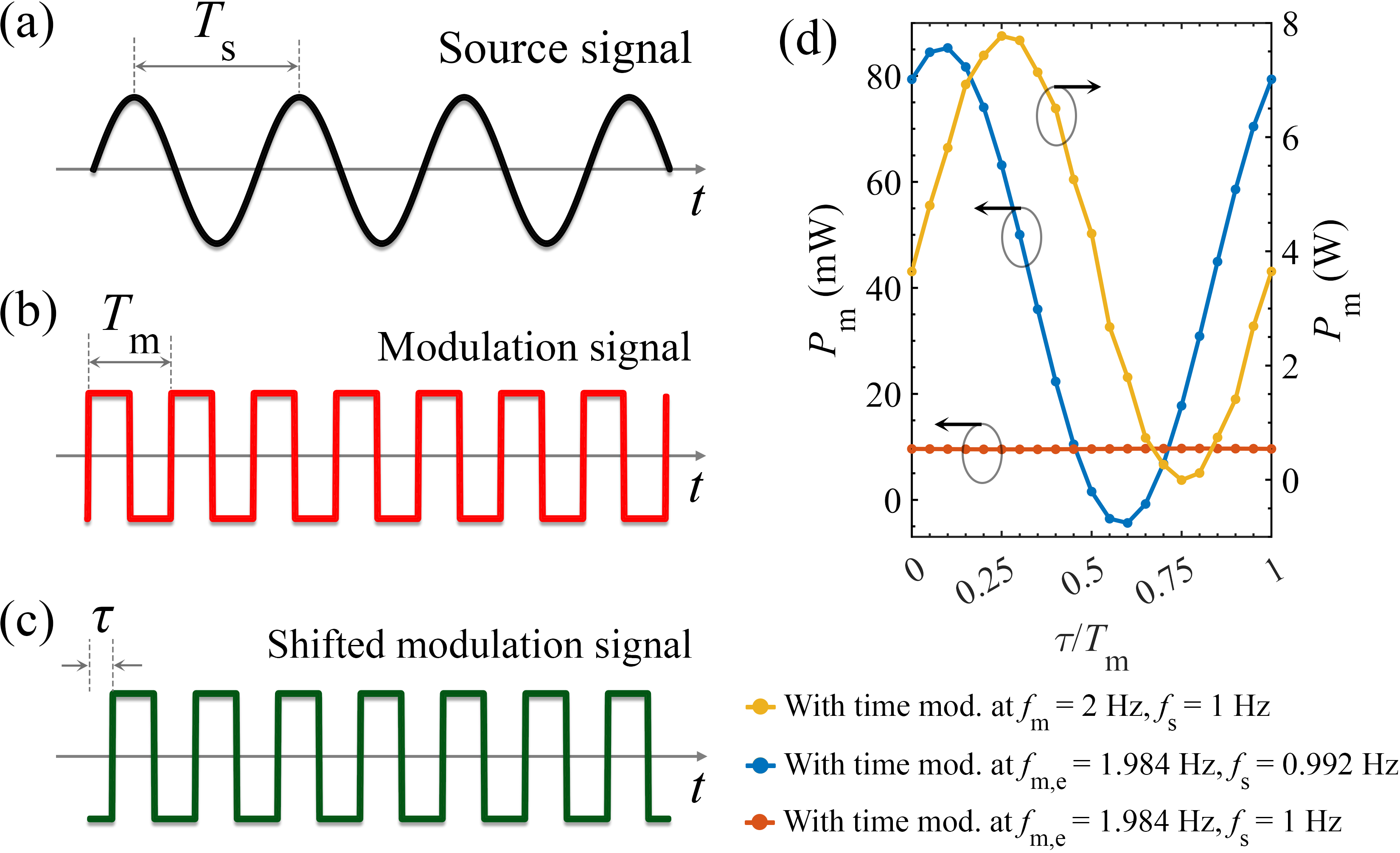}\caption{(a) Source sinusoidal signal with a period of $T_{\mathrm{s}}=2T_{\mathrm{m}}$.
		(b) Piece-wise constant time modulated damper with a period of $T_{\mathrm{m}}$
		and (c) shifted by a delay $\tau$. (d) Time-averaged power $\mathrm{\mathit{P}_{m}}$
		delivered by the time-varying damper versus delay $\tau$ for three
		different scenarios. The value of $\mathrm{\mathit{P}_{m}}$ at EPD
		modulation frequency $f_{\mathrm{m,e}}=1.984\:\mathrm{Hz}$ goes to
		a small value of $\mathrm{\mathit{P}_{m}}=0.04\:\mathrm{mW}$ at the
		selected delay of $\tau/T_{\mathrm{m}}=0.51$.\label{Fig: Signals}}
\end{figure}

According to Eq. (\ref{eq: GenEigVec}), degenerate eigenvalues result
in degenerate eigenvectors. At an EPD, the transition matrix is similar
to a Jordan block with two degenerate eigenvalues $\lambda_{1}=\lambda_{2}$
that are associated with degenerate eigenvectors. Thus, degenerate
eigenvalues are sufficient to guarantee the degeneracy of the eigenvectors
and hence the occurrence of an EPD. The formulation presented here
is general and can be applied to any system described by the differential
equations of Eq. (\ref{eq:Damper_Main}) in the absence of an external
source term. As a matter of energy analysis, a time-periodic system
is not isolated and it is continually interacting with the source
of time-modulation. Therefore, energy can be transferred into or out
of the system, i.e., the system can gain or lose energy from the time-variation
source.

\begin{figure*}[t]
	\centering{}\includegraphics[width=0.8\textwidth]{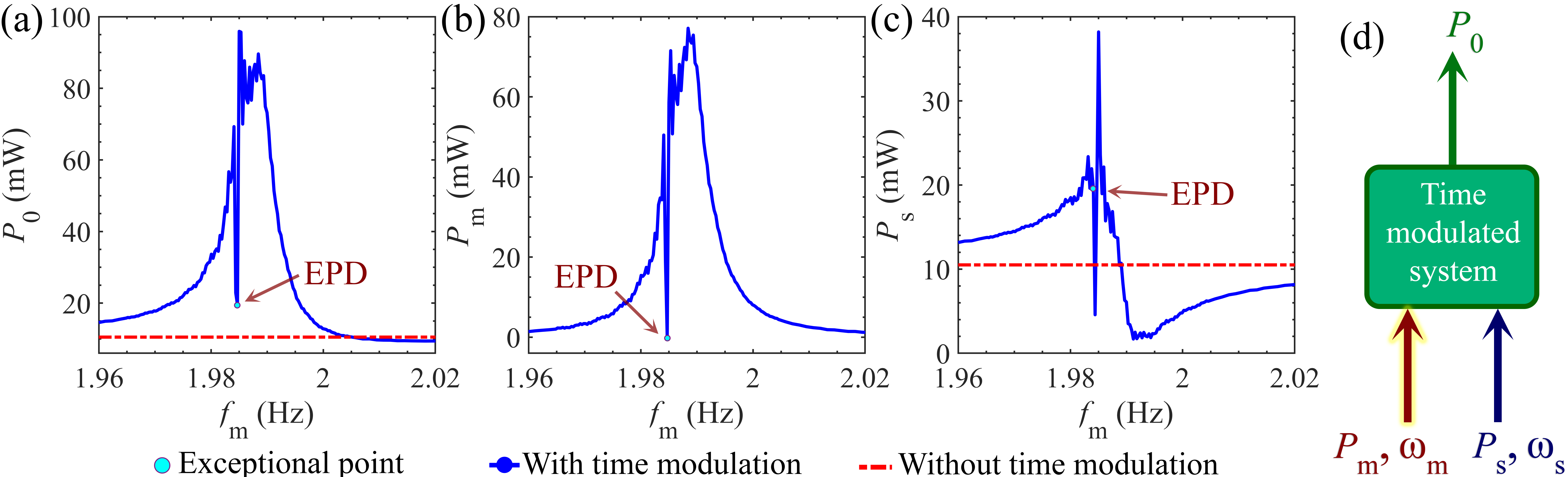}\caption{Time-averaged power levels, after reaching steady state, as in Fig.
		\ref{Fig:modulation_fm}, but assuming a time delay $\tau=0.51T_{\mathrm{m}}$
		in the modulation of the damper, for the case of $f_{\mathrm{s}}=f_{\mathrm{m,e}}/2$.\label{Fig: excitation_tau_minPp}}
\end{figure*}

The eigenfrequency dispersion diagram of the mechanical system with
the mentioned parameters is shown in Figs. \ref{Fig: Dispersion_diagram}(a)
and (b). EPD happens at two modulated frequencies of $f_{\mathrm{m,e}}=1.984\:\mathrm{Hz}$
and $f_{\mathrm{m,e}}=2.015\:\mathrm{Hz}$. Note that the quality
factor of a resonating system is $Q=\mathrm{Re}\left(\omega\right)/\left(2\mathrm{Im}\left(\omega\right)\right)$,
which is higher for smaller $\mathrm{Im}\left(\omega\right)$. Usually,
a higher quality factor implies a higher power harvested by the system.
A modulation gap in the dispersion diagram of eigenfrequencies happens
when the two eigenfrequencies have two non-vanishing imaginary parts,
i.e., between two closeby EPDs. When the modulation frequency is selected
in the middle of the modulation gap, i.e., $f_{\mathrm{m}}=2\:\mathrm{Hz}$,
one eigenfrequency has the smaller $\mathrm{Im}\left(\omega\right)$,
corresponding to a better quality factor compared to the two neighbor
EPDs at slightly higher and lower frequencies. Thus, we expect the
largest improvement in harvested power in the middle of the modulation
gap. Also, by selecting $f_{\mathrm{m,e}}=1.984\:\mathrm{Hz}$ the
system experiences harsh changes around $f_{\mathrm{s}}=f_{\mathrm{m,e}}/2=0.992\:\mathrm{Hz}$
due to the degeneracy of the eigenfrequencies. The spectrum of the
displacement $z\left(t\right)$ when the source-free system is modulated
at $f_{\mathrm{m}}=f_{\mathrm{m,e}}=1.984\:\mathrm{Hz}$ and it is
excited by an initial condition (see Appendix A for
the calculation details) is illustrated in Fig. \ref{Fig: Dispersion_diagram}(c).
The spectrum peak can be observed at $f_{\mathrm{e}}=f_{\mathrm{m,e}}/2=0.992\:\mathrm{Hz}$,
that is same as the one obtained from solving the eigenvalue problem
shown in Figs. \ref{Fig: Dispersion_diagram}(a) and (b).

When $f_{\mathrm{m}}=f_{\mathrm{m,e}}=1.984\:\mathrm{Hz}$, the mechanical
system operates at the EPD as shown in Figs. \ref{Fig: Dispersion_diagram}(a)
and (b). The frequency domain spectrum in Fig. \ref{Fig: Dispersion_diagram}(c)
shows that the first harmonic at $f=f_{\mathrm{e}}=f_{\mathrm{m,e}}/2=0.992\:\mathrm{Hz}$,
carries the maximum power. After setting $f_{\mathrm{s}}=f_{\mathrm{e}}=f_{\mathrm{m,e}}/2=0.992\:\mathrm{Hz}$,
Figs. \ref{Fig:modulation_fm}(a)-(c) show the powers by varying the
modulation frequency around its EPD value $f_{\mathrm{m,e}}=1.984\:\mathrm{Hz}$.
In this plot, time-modulated case is in blue curve, whereas the red
dashed-line reminds the power values of the system without time-modulation.
The numerical results show that the system operating very near to
EPD (the EPD is the cyan point) harvests more power from the external
source ($\mathrm{\mathit{P}_{s}}=20.9\:\mathrm{mW}$) compared to
the system without time-modulation ($\mathrm{\mathit{P}_{s}}=10.5\:\mathrm{mW}$).
Thus, time modulation leads to harvest more power from external vibration,
with an improvement of $99\%$. In addition, the time-modulated element
delivers the power of $\mathrm{\mathit{P}_{m}}=80.3\:\mathrm{mW}$
to the system and the constant part of the damper absorbs $\mathrm{\mathit{P}_{0}}=\mathrm{\mathit{P}_{s}}+\mathrm{\mathit{P}_{m}}=101.2\:\mathrm{mW}$
from the system, at the EPD frequency.

The eigenfrequencies near the EPD are very sensitive to a system's
variation, like a small change in the modulation frequency, as discussed
in Appendix C. When such a small relative perturbation
$\delta_{\mathrm{m}}=\left(f_{\mathrm{m}}-f_{\mathrm{m,e}}\right)/f_{\mathrm{m,e}}$
is applied, the resulting two distinct eigenfrequencies $f_{\mathrm{1,2}}\left(\delta_{\mathrm{m}}\right)$
are estimated using the Puiseux series power expansion $f_{\mathrm{1,2}}\left(\delta_{\mathrm{m}}\right)\approx f_{\mathrm{e}}\mp j\left(f_{\mathrm{m}}/2\pi\right)\alpha_{1}\sqrt{\delta_{\mathrm{m}}}$
\cite{kazemi2022experimental}, where $\alpha_{1}$ is the first-order
coefficient of the series. The square root function demonstrates that
the eigenfrequencies are highly sensitive to modulation frequency
perturbations around the EPD, as shown in Fig. \ref{Fig: Dispersion_diagram}.
This is reflected by the power levels that change dramatically when
a small change in $f_{\mathrm{m}}$ is applied, as shown in Figs.
\ref{Fig:modulation_fm}(a)-(c). To harvest more power, the modulation
frequency must be close to the EPD frequency, but the modulation frequency
value must also be precisely chosen.

To control the power provided by the time modulated portion of the
circuit, Fig. \ref{Fig: Signals}(d) shows the time-averaged power
$P_{\mathrm{m}}$ delivered by $c_{\mathrm{m}}\left(t\right)$ versus
delay, for $0<\tau<T_{\mathrm{m}}$. Three combinations of modulation
and source frequencies are considered. For the two cases shown with
blue and yellow curves in Fig. \ref{Fig: Signals}(d), we assume that
the frequency of the source is fixed to $f_{\mathrm{s}}=f_{\mathrm{m}}/2$,
hence, the period of the modulation signal is half of the source's
one when operating at the EPD ($T_{\mathrm{m,e}}=T_{\mathrm{s}}/2$).
In the third case (orange line), we assume that the modulation frequency
is the EPD one, $f_{\mathrm{m}}=f_{\mathrm{m,e}}$, and source frequency
at $f_{\mathrm{s}}=1\:\mathrm{Hz}$. It is clear that the variation
of the delay $\tau$ has a strong effect on $P_{\mathrm{m}}$ when
the modulation frequency is selected as $f_{\mathrm{m}}=2f_{\mathrm{s}}$
(both blue and yellow curves). However, in the case shown by the orange
curve, which is a slight modification from the other two cases, $P_{\mathrm{m}}$
is more or less constant and the delay does not have much effect on
it. When $f_{\mathrm{m}}=f_{\mathrm{m,e}}$, and $f_{\mathrm{s}}=f_{\mathrm{m}}/2$
(blue curve), and at a specific delay $\tau=0.51T_{\mathrm{m}}$,
the power delivered by the time-varying damper reaches the minimum
and it is as small as $\mathrm{\mathit{P}_{m}}=0.04\:\mathrm{mW}$.
By assuming the latter particular condition with $f_{\mathrm{s}}=f_{\mathrm{m,e}}/2$
and $\tau=0.51T_{\mathrm{m}}$, the numerical results in Fig. \ref{Fig: excitation_tau_minPp}
show the source, modulation and constant damper powers by varying
the modulation frequency (compared to the results in Fig. \ref{Fig:modulation_fm},
where $\tau=0).$ Blue curves represent the power in the system with
time modulation (note that the modulated power is near zero at the
EPD), while red curves is the reminder of the power levels for the
case without time modulation, as was done in Fig. \ref{Fig:modulation_fm}.
At the EPD modulation frequency $f_{\mathrm{m}}=f_{\mathrm{m,e}}=1.984\:\mathrm{Hz}$
(cyan point), the power extracted from the source vibration is $\mathrm{\mathit{P}_{s}}=19.38\:\mathrm{mW}$
and the power delivered to the constant part of the damper $\mathrm{\mathit{c}_{0}}$
is $\mathrm{\mathit{P}_{0}}=19.42\:\mathrm{mW}$. Thus, time modulation
improves power harvesting by $85\%$ compared to the case without
time modulation. Note also that in this case the power delivered by
the modulation $\mathrm{\mathit{P}_{m}}=0.04\:\mathrm{mW}$ is very
small; so $\mathrm{\mathit{P}_{0}}\approx\mathrm{\mathit{P}_{s}}$.
However, the calculated powers around the EPD vary dramatically when
changing the modulation frequency. That is one of the most peculiar
properties associated with an EPD, as already depicted in Figs. \ref{Fig:modulation_fm},
and \ref{Fig: excitation_tau_minPp}. The reason is that the eigenstates
at the EPD are extremely sensitive to any perturbation, as shown in
Fig. \ref{Fig: Dispersion_diagram}, which causes a large degree of
variation when interacting with forced excitation. The system could
possibly harvest even more power with a precise choice of modulation
frequency close to EPD and proper time delay. However, even without
a precise modulation frequency choice, the system harvests more power
on average, over $f_{\mathrm{s}}$ or $f_{\mathrm{m}}$ variation.

As a conclusion, we have shown that a mechanical resonator with time
modulation harvests much higher power (even ten times higher) from
ambient vibration than its counterpart without time modulation. This
effect can make a difference in applications where the ambient source
amplitude is very low. Moreover, using the concept of second-order
EPDs, we also explain the existence of a very sharp spectral peak
at some modulation frequencies. The power levels vary rapidly with
only a very slight variation in a parameter (like a modulation frequency)
so great care must be taken when selecting the values of the modulation
signal frequency. Indeed, using the Puiseux fractional series expansion,
we have demonstrated that the degenerated system's eigenfrequency
is highly sensitive to perturbations in the modulation frequency,
and the physics associated with an EPD in an LTP mechanical system
is vital to understand the EPD-related very narrow frequency feature
of the power transfer mechanism. This effect could be exploited for
sensing applications or for conceiving a very precise spectrometer.

\section*{Acknowledgments}
This material is based upon work supported by the Air Force Office
of Scientific Research (AFOSR) under Grant No. FA9550-19-1-0103.

\section*{APPENDIX}
\subsection{Dual Circuit with Time-modulated Conductance}

\textcolor{black}{We show }the analogous (dual) system made of a LC
resonator and a linear time-periodic (LTP) conductance connected to
the external source in series to the capacitor as shown in Fig. \ref{Fig: Dual_circuit}.
In order to calculate the power in the dual LTP system, we consider
piece-wise constant time-periodic conductance $G\left(t\right)$,
i.e., $G\left(t\right)=G_{0}+G_{\mathrm{m}}$ during the time interval
$0\leq t<T_{\mathrm{m}}/2$, and $G\left(t\right)=G_{0}-G_{\mathrm{m}}$
during the time interval $T_{\mathrm{m}}/2\leq t<T_{\mathrm{m}}$.
Analogously to the mechanical LTP system, we define the system state
vector as $\boldsymbol{\Psi}\left(t\right)=\left[v\left(t\right),\dot{v}\left(t\right)\right]^{\mathrm{T}}$,
where $v\left(t\right)$ is the voltage on the inductor and $\dot{v}\left(t\right)$
is its time derivative. In general, $G\left(t\right)$ can be either
lossy or gain (positive or negative respectively). Kirchhoff's circuit
laws apply to time-varying circuits as follows:

\begin{figure}[t]
	\centering{}\includegraphics[width=0.4\columnwidth]{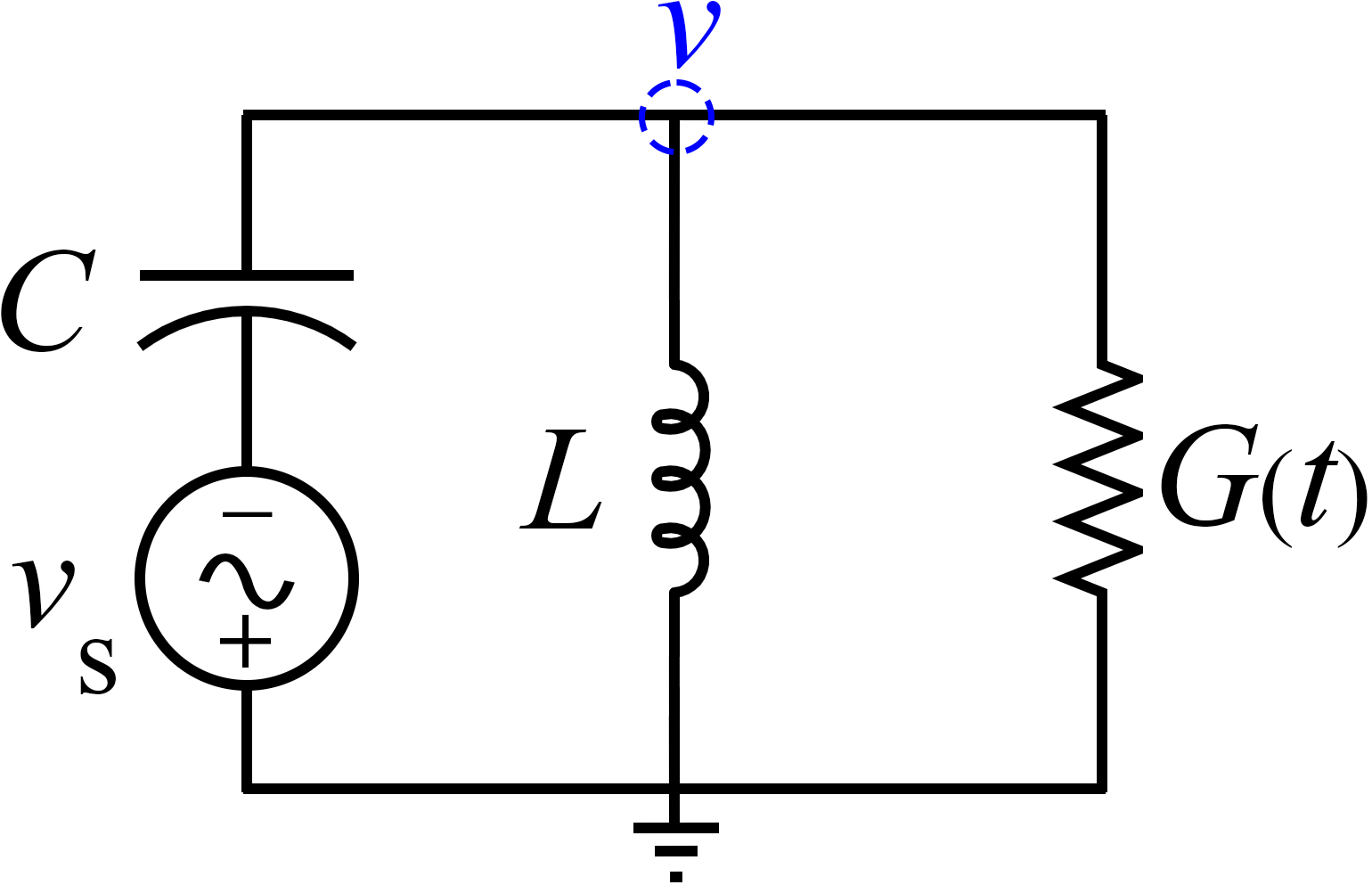}\caption{The dual time-varying circuit where an inductance is connected in
		parallel to a time-varying conductance, with an external source $v_{\mathrm{s}}\left(t\right)$
		in series to the capacitor. The conductance is designed to have a
		two-level piece-wise constant time-periodic conductance, where $G\left(t\right)=G_{0}+G_{\mathrm{m}}$
		during the time interval $0\protect\leq t<T_{\mathrm{m}}/2$, and
		$G\left(t\right)=G_{0}-G_{\mathrm{m}}$ during the time interval $T_{\mathrm{m}}/2\protect\leq t<T_{\mathrm{m}}$.\label{Fig: Dual_circuit}}
\end{figure}

\begin{equation}
	G\left(t\right)\dot{v}+\frac{v}{L}+C\left(\ddot{v}+\ddot{v}_{\mathrm{s}}\right)=0.
\end{equation}

The circuit equation is rewritten as

\begin{equation}
	\ddot{v}+2\alpha\left(t\right)\dot{v}+\omega_{0}^{2}v=-\ddot{v}_{\mathrm{s}}.
\end{equation}
where $\alpha\left(t\right)=\alpha_{0}+\alpha_{\mathrm{m}}\left(t\right)=G_{0}/\left(2C\right)+G_{\mathrm{m}}\left(t\right)/\left(2C\right)$
is the time-modulated damping factor and $\omega_{0}=1/\sqrt{LC}$
is the natural frequency of the unmodulated and lossless circuit.
Assuming time harmonic dependence of the form $v\propto e^{j\omega t}$
for the unmodulated homogeneous circuit, we obtain the eigenfrequencies
as $\omega=j\alpha(1\pm\sqrt{1-\omega_{0}^{2}/\alpha^{2}})$. By writing
the differential equation in the eigenvalue problem format, the time
evolution of the state vector $\boldsymbol{\Psi}\left(t\right)$ is
given by

\begin{equation}
	\begin{array}{c}
		d\boldsymbol{\Psi}\left(\mathit{t}\right)/d\mathit{t}=\mathbf{\underline{M}}_{\mathrm{c}}\left(\mathit{t}\right)\boldsymbol{\Psi}\left(\mathit{t}\right)+\left(\begin{array}{c}
			0\\
			-\ddot{v}_{s}
		\end{array}\right),\\
		\\
		\mathbf{\underline{M}}_{\mathrm{c}}\left(\mathit{t}\right)=\left(\begin{array}{cc}
			0 & 1\\
			-\omega_{0}^{2} & -2\alpha\left(\mathit{t}\right)
		\end{array}\right),
	\end{array}\label{eq:GLC_Eigenvalue_P}
\end{equation}
where $\mathbf{\underline{M}}_{\mathrm{c}}\left(\mathit{t}\right)$
is the equivalent circuit matrix. The differential equation and the
circuit matrix are dual to the time-varying mechanical system whose
time-varying damper is connected to the spring and mass. The duality
transformation is $k\rightarrow1/L$, $m\rightarrow C$ and $c\left(\mathit{t}\right)\rightarrow G\left(\mathit{t}\right)$
and both systems have an external excitation $\ddot{y}(t)\rightarrow\ddot{v}_{\mathrm{s}}(t)$
as summarized in Table \ref{tab:Component-values} \cite{le1952duality}.
Also, we show the duality of the characteristic equations in mechanical
systems and their dual version in electric circuits in Table \ref{tab:Equations}.
By applying the conversion between force and current ($\mathscr{F}\longleftrightarrow i$)
and velocity and voltage ($\dot{z}\longleftrightarrow v$), Newton's
equations and instantaneous mechanical power relate to the electric
dual equations. We analyze the dual LTP circuit by using the Keysight
Advanced Design System (ADS) time-domain simulator to calculate the
power. We excite the circuit with a sinusoidal source at a frequency
of $f_{\mathrm{s}}$ and the amplitude of $10\:\mathrm{mV}$. Also,
the capacitor has a $10\:\mathrm{mV}$ as an initial condition for
the case where no external source excites the system. We numerically
calculate the power using a built-in power block in the simulator
and then report the time-averaged power based on $1000$ time periods
after the time domain signal saturates (i.e., for the time window
from $3500\:\mathrm{s}$ to $4500\:\mathrm{s}$).

\begin{table*}[tbh]
	\centering{}\caption{Component values in the mechanical system and their dual values in
		the dual electrical circuit.\label{tab:Component-values}}
	\begin{tabular}{|c|c|c|}
		\hline 
		Mechanical system & Dual electrical circuit & Duality\tabularnewline
		\hline 
		\hline 
		$k=4\pi^{2}\:\mathrm{N/m}$ & $L=0.025\:\mathrm{H}$ & $k\rightarrow1/L$\tabularnewline
		\hline 
		$m=1\:\mathrm{kg}$ & $C=1\:\mathrm{F}$ & $m\rightarrow C$\tabularnewline
		\hline 
		$c_{0}=0.1\:\mathrm{Ns/m}$ & $G_{0}=0.1\:\mathrm{S}$ & $c_{0}\rightarrow G_{0}$\tabularnewline
		\hline 
		$c_{\mathrm{m}}=0.15\:\mathrm{Ns/m}$ & $G_{\mathrm{m}}=0.15\:\mathrm{S}$ & $c_{\mathrm{m}}\rightarrow G_{\mathrm{m}}$\tabularnewline
		\hline 
		$c\left(t\right)=\left\{ \begin{array}{c}
			c_{0}+c_{\mathrm{m}},\:\:\:0\leq t<T_{\mathrm{m}}/2\\
			c_{0}-c_{\mathrm{m}},\:\:\:T_{\mathrm{m}}/2\leq t<T_{\mathrm{m}}
		\end{array}\right.$ & $G\left(t\right)=\left\{ \begin{array}{c}
			G_{0}+G_{\mathrm{m}},\:\:\:0\leq t<T_{\mathrm{m}}/2\\
			G_{0}-G_{\mathrm{m}},\:\:\:T_{\mathrm{m}}/2\leq t<T_{\mathrm{m}}
		\end{array}\right.$ & $c\left(t\right)\rightarrow G\left(t\right)$\tabularnewline
		\hline 
	\end{tabular}
\end{table*}

\begin{table*}[tbh]
	\centering{}\caption{Dual equations in the mechanical system and dual electrical circuit,
		where $\mathscr{F}$ is the force and $i$ is the current.\label{tab:Equations}}
	\begin{tabular}{|c|c|c|c|}
		\hline 
		\multicolumn{2}{|c|}{Mechanical system} & \multicolumn{2}{c|}{Dual electrical circuit}\tabularnewline
		\hline 
		\hline 
		Spring & $\mathscr{F}=kz$ & $i=(1/L)\int v\:dt^{'}$ & Inductor\tabularnewline
		\hline 
		Mass & $\mathscr{F}=m\ddot{z}$ & $i=C\dot{v}$ & Capacitor\tabularnewline
		\hline 
		Damper & $\mathscr{F}=c\dot{z}$ & $i=Gv$ & Conductance\tabularnewline
		\hline 
		Mechanical Power & $p=\mathscr{F}\dot{z}$ & $p=iv$ & Electrical Power\tabularnewline
		\hline 
		\multicolumn{4}{|c|}{Duality}\tabularnewline
		\hline 
		\multicolumn{4}{|c|}{$\mathscr{F}\longleftrightarrow i$}\tabularnewline
		\hline 
		\multicolumn{4}{|c|}{$\dot{z}\longleftrightarrow v$}\tabularnewline
		\hline 
	\end{tabular}
\end{table*}

\subsection{Vibration Conversion}
As already mentioned in the paper, $c_{0}=c_{\mathrm{p}}+c_{\mathrm{t}}$
is responsible for the energy losses within the system due to parasitic
loss mechanisms $c_{\mathrm{p}}$ (e.g., viscous friction with air),
and by the intentional mechanism of damping $c_{\mathrm{t}}$, i.e.,
the mechanical energy extracted by the transduction mechanism. This
model is based on the idea that converting energy from an oscillating
mass to electricity (whatever the mechanism is) can be modeled as
a linear damper in a mass-spring system. This model is quite accurate
for certain types of electromechanical converters, such as those analyzed
by Williams and Yates \cite{williams1996analysis}. For other types
of converter, such as electrostatic and piezoelectric, the model may
be modified. However, the conversion will always result in a loss
of mechanical kinetic energy, which can be referred to as damping
\cite{roundy2003study}. Despite the fact that the current damper
model does not accurately model all kinds converter types, the present
analysis can be extended to electrostatic and piezoelectric systems
\cite{roundy2003study}. The power extracted from the mechanical system
via $c_{\mathrm{t}}$ is due to electrically induced damping and it
is constitutes the whole time-averaged power $P_{0}$ if the parasitic
damping vanishes. The instantaneous power in the constant part of
the damper $c_{0}$, i.e., the combination of parasitic loss mechanisms
$c_{\mathrm{p}}$ and transduction mechanism $c_{\mathrm{t}}$, can
be calculated as a product of induced force $c_{0}\dot{z}$ and velocity
$\dot{z}$. Thus, the absorbed instantaneous power in the constant
part of the damper is expressed by

\begin{equation}
	p_{\mathrm{0}}(t)=c_{\mathrm{0}}\dot{z}^{2}.
\end{equation}

The total time-averaged power $P_{0}$ is calculated by averaging
the time domain expression. In a monochromatic regime, assuming no
modulation, the total time-averaged power dissipated within the damper,
i.e., the power extracted via the transduction mechanism and the power
lost by parasitic damping mechanisms, is given by \cite{beeby2007micro,spies2015handbook,wei2017comprehensive}

\begin{equation}
	P_{0}=\frac{m\zeta_{\mathrm{0}}\omega_{0}\omega_{\mathrm{s}}^{2}\left(\frac{\omega_{\mathrm{s}}}{\omega_{0}}\right)^{3}y_{0}^{2}}{\left(2\zeta_{\mathrm{0}}\frac{\omega_{\mathrm{s}}}{\omega_{0}}\right)^{2}+\left(1-\left(\frac{\omega_{\mathrm{s}}}{\omega_{0}}\right)^{2}\right)^{2}},
\end{equation}
where $y_{0}$ is the magnitude of the source vibration, $\zeta_{0}=\zeta_{\mathrm{t}}+\zeta_{\mathrm{p}}=c_{0}/\left(2m\omega_{0}\right)$
is the constant damping ratio. Maximum power dissipation within the
generator occurs when the device is operated at $\omega_{\mathrm{s}}=\omega_{0}$,
and in this case the total time-averaged power dissipated in the constant
part of the damper is given by

\begin{equation}
	P_{0}=\frac{m\omega_{0}^{3}y_{0}^{2}}{4\zeta_{\mathrm{0}}}.\label{eq:Total Power}
\end{equation}

\subsection{Sensitivity to Perturbation}
Sensitivity of a system\textquoteright s observable to a particular
parameter is a measure of how much a perturbation to that parameter
affects the observable quantity of the system. The eigenvalues of
the system at exceptional points of degeneracy (EPDs) are extremely
sensitive to parameter changes, which is a significant feature. Applying
a perturbation to a system parameter such as the modulation frequency
$\delta_{\mathrm{m}}=\left(f_{\mathrm{m}}-f_{\mathrm{m,e}}\right)/f_{\mathrm{m,e}}$,
leads to a perturbed transition matrix $\underline{\boldsymbol{\Phi}}\left(\delta_{\mathrm{m}}\right)$
and perturbed eigenvalues $\lambda_{p}\left(\delta_{\mathrm{m}}\right)$,
with $p=1,2$. Therefore, the degenerate resonance frequency occurring
at the EPD $f_{\mathrm{e}}$, splits into two distinct resonance frequencies
$f_{p}\left(\delta_{\mathrm{m}}\right)$, due to a small perturbation
$\delta_{\mathrm{m}}$. We can calculate the perturbed eigenvalues
near the EPD by using the convergent Puiseux fractional power series
expansion, with coefficients calculated using the explicit recursive
formulas in \cite{welters2011explicit}. In the presented mechanical
system with a second-order EPD, we utilize a first-order approximation
of the perturbed eigenvalues as

\begin{equation}
	\lambda_{p}\left(\delta_{\mathrm{m}}\right)\approx\lambda_{\mathrm{e}}+\left(-1\right)^{p}\alpha_{1}\sqrt{\delta_{\mathrm{m}}},
\end{equation}
where $\lambda_{\mathrm{e}}$ is the eigenvalue at EPD and the first
order coefficient is expressed by

\begin{equation}
	\alpha_{1}=\left.\left(-\frac{\partial H\left(\delta_{\mathrm{m}},\lambda\right)/\partial\delta_{\mathrm{m}}}{\frac{1}{2!}\partial^{2}H\left(\delta_{\mathrm{m}},\lambda\right)/\partial\lambda^{2}}\right)^{\frac{1}{2}}\right|_{\delta_{\mathrm{m}}=0,\,\lambda=\lambda_{\mathrm{e}}},\label{eq:Puiseux}
\end{equation}
where $H\left(\delta_{\mathrm{m}},\lambda\right)=\det\left(\underline{\boldsymbol{\Phi}}\left(\delta_{\mathrm{m}}\right)-\lambda\underline{\boldsymbol{\mathrm{I}}}\right)$
and $\underline{\boldsymbol{\mathrm{I}}}$ is the $2\times2$ identity
matrix. The perturbed resonance frequencies are approximately calculated
as

\begin{equation}
	f_{p}\left(\delta_{\mathrm{m}}\right)\approx f_{\mathrm{e}}\pm j\frac{f_{\mathrm{m}}}{2\pi}\left(-1\right)^{p}\alpha_{1}\sqrt{\delta_{\mathrm{m}}}.
\end{equation}

This formula proves that the time-modulated system supporting the
EPD is very sensitive to variations in the modulation frequency $f_{\mathrm{m}}$.
Figures 2, 4 and 6 of the paper demonstrate that the harvested power
is very sensitive to variations in the system's parameters when operating
at or near an EPD.


\end{document}